\documentclass{article}
\usepackage{spconf,amsmath,amssymb,graphicx,url}
\usepackage{tikz}
\usepackage{comment}
\usepackage[
    backend=biber,
    style=ieee,
    maxbibnames=3,
    maxcitenames=3,
    doi=false,isbn=false,url=false,eprint=false
]{biblatex} 

\newcommand{\Fig}[1]{Figure~\ref{fig:#1}} 
\newcommand{\Eq}[1]{Eq.~(\ref{eq:#1})} 
\renewcommand{\Vec}[1]{\textrm{\boldmath $#1$}} 

\newcommand{\pt}[1]{\left(#1\right)} 
\newcommand{\br}[1]{\left[#1\right]} 

\newcommand{\Normal}[1]{\mathcal{N}\pt{#1}}

\newcommand{\drawfig}[4]{ 
    \begin{figure}[#1]
    \centering 
    \vspace{0mm}
    \includegraphics[width=#2]{#3} 
    \vspace{-3.5mm}
    \caption{#4}
    \label{fig:#3}
    \vspace{-4mm}
    \end{figure}
}

\DeclareSourcemap{
	\maps[datatype=bibtex, overwrite=true]{
		\map{
			\step[fieldsource=booktitle,
			match=\regexp{.*Interspeech.*},
			replace={Proc. Interspeech}]
			\step[fieldsource=journal,
			match=\regexp{.*INTERSPEECH.*},
			replace={Proc. Interspeech}]
			\step[fieldsource=booktitle,
			match=\regexp{.*icassp_inpress.*},
			replace={ICASSP (in press)}]
			\step[fieldsource=booktitle,
			match=\regexp{.*International.*Conference.*on.*Acoustics,.*Speech.*and.*Signal.*Processing.*},
			replace={Proc. ICASSP}]
			\step[fieldsource=booktitle,
			match=\regexp{.*International.*Conference.*on.*Learning.*Representations.*},
			replace={ICLR}]
			\step[fieldsource=booktitle,
			match=\regexp{.*International.*Conference.*on.*Machine.*Learning.*},
			replace={ICML}]
			\step[fieldsource=booktitle,
			match=\regexp{.*Automatic.*Speech.*Recognition.*and.*Understanding.*},
			replace={Proc. ASRU}]
			\step[fieldsource=booktitle,
			match=\regexp{.*Spoken.*Language.*Technology.*},
			replace={Proc. SLT}]
			\step[fieldsource=booktitle,
			match=\regexp{.*Speech.*Synthesis.*Workshop.*},
			replace={Proc. SSW}]
			\step[fieldsource=booktitle,
			match=\regexp{.*workshop.*on.*speech.*synthesis.*},
			replace={Proc. SSW}]
			\step[fieldsource=booktitle,
			match=\regexp{.*Advances.*in.*neural.*information.*processing.*},
			replace={Proc. NIPS}]
			\step[fieldsource=booktitle,
			match=\regexp{.*Advances.*in.*Neural.*Information.*Processing.*},
			replace={Proc. NIPS}]
			\step[fieldsource=booktitle,
			match=\regexp{.*Workshop.*on.* Applications.* of.* Signal.*Processing.*to.*Audio.*and.*Acoustics.*},
			replace={Proc. WASPAA}]
			\step[fieldsource=booktitle,
			match=\regexp{.*International.*Conference.*on.*Language.*Resources.*and.*Evaluation.*},
			replace={Proc. LREC}]
			\step[fieldsource=journal,
			match=\regexp{.*Spontaneous.*Speech.*Processing.*and.*Recognition},
			replace={Proc. SSPR}]
			\step[fieldsource=publisher,
			match=\regexp{.+},
			replace={{}}]
			\step[fieldsource=month,
			match=\regexp{.+},
			replace={{}}]
			\step[fieldsource=location,
			match=\regexp{.+},
			replace={{}}]
			\step[fieldsource=address,
			match=\regexp{.+},
			replace={{}}]
			\step[fieldsource=organization,
			match=\regexp{.+},
			replace={{}}]
			\step[fieldsource=doi,
			match=\regexp{.+},
			replace={{}}]
			\step[fieldsource=url,
			match=\regexp{.+},
			replace={{}}]
			\step[fieldsource=editor,
			match=\regexp{.+},
			replace={{}}]
		}
	}
} 

\title{Mid-attribute Speaker Generation using Optimal-Transport-based Interpolation of Gaussian Mixture Models}
\name{Aya Watanabe, Shinnosuke Takamichi, Yuki Saito, Detai Xin, Hiroshi Saruwatari\thanks{This work is supported by JSPS KAKENHI 21H04900 (practical experiment) and Moonshot R\&D Grant Number JPMJPS2011 (algorithm development). We also appreciate Takaaki Saeki and Yuta Matsunaga of the University of Tokyo for their help.}}
\address{The University of Tokyo, Japan.}

\begin{document}
\ninept
\maketitle
\setlength{\tabcolsep}{1mm} 
\setlength{\abovedisplayskip}{3pt} 
\setlength{\belowdisplayskip}{3pt} 

\begin{abstract}
    In this paper, we propose a method for intermediating multiple speakers' attributes and diversifying their voice characteristics in ``speaker generation,'' an emerging task that aims to synthesize a nonexistent speaker's naturally sounding voice.
    The conventional TacoSpawn-based speaker generation method represents the distributions of speaker embeddings by Gaussian mixture models (GMMs) conditioned with speaker attributes.
    Although this method enables the sampling of various speakers from the speaker-attribute-aware GMMs, it is not yet clear whether the learned distributions can represent speakers with an intermediate attribute (i.e., mid-attribute).
    To this end, we propose an optimal-transport-based method that interpolates the learned GMMs to generate nonexistent speakers with mid-attribute (e.g., gender-neutral) voices.
    We empirically validate our method and evaluate the naturalness of synthetic speech and the controllability of two speaker attributes: gender and language fluency. The evaluation results show that our method can control the generated speakers' attributes by a continuous scalar value without statistically significant degradation of speech naturalness.
\end{abstract} \vspace{-1mm}

\begin{keywords} 
    speech synthesis, cross-lingual speech synthesis, multi-speaker speech synthesis, speaker generation
\end{keywords}

\vspace{-2mm}
\section{Introduction} \label{sec:introduction}
\vspace{-2mm}
Despite the improved quality of synthetic speech through deep neural network (DNN)-based text-to-speech (TTS)~\cite{oord2016wavenet,wang2017tacotron,ren2020fastspeech}, diversifying speakers' voices remains challenging. One approach to increase speaker diversity is multi-speaker TTS~\cite{gibiansky2017deep}, in which a single TTS model reproduces the voice characteristics of speakers included in a multi-speaker corpus. However, the training requires sufficient speech data for each speaker to achieve high-quality TTS. Although few-shot speaker adaptation~\cite{arik2018neural, jia2018transfer, moss2020boffin, chen2021adaspeech} and zero-shot speaker encoding~\cite{jia2018transfer, choi2020attentron, casanova2021sc, cooper2020zero, chien2021investigating} can reproduce a target speaker's voice using only a few utterances of the speaker, they still need an existent speaker's speech data. Some work has attempted to generate nonexistent speakers from a trained multi-speaker TTS model to deal with the difficulty in collecting speech data of existent speakers~\cite{jia2018transfer, mitsui2021deep, valle2021flowtron}. Recently, Stanton et al.~\cite{stanton2022speaker} define this task as ``speaker generation,'' where the purpose is to synthesize nonexistent speakers' natural-sounding voices and achieve practical applications such as audiobook readers and video production.

Stanton et al.~\cite{stanton2022speaker} proposed TacoSpawn as a method for resolving the speaker generation. TacoSpawn jointly learns two DNNs: a multi-speaker TTS model and an encoder that defines the parametric distributions of speaker embeddings. The former generates a target speaker's mel-spectrogram from the input text and the speaker embedding. The latter learns the distributions of speaker embeddings as Gaussian mixture models (GMMs) for each ``speaker attribute'' (or ``speaker metadata''~\cite{stanton2022speaker}) representing the attributes (e.g., gender) of a specific speaker. The combination of these two models achieves TTS of not only existent speakers' voice, but also nonexistent ones' by sampling new embeddings from the speaker-attribute-aware GMM.

Learned parametric distributions of speaker embeddings by the conventional TacoSpawn method can potentially synthesize more diverse speakers' voices. For example, we can transform or interpolate the speaker embedding distributions to define a new distribution for speaker generation. In other words, at present, TacoSpawn only handles speakers with categorical attributes. However, by combining the distributions of individual attributes, it is possible to handle speakers of non-categorical attributes.
Such ``mid-attribute speaker generation'' method would extend the application range of TTS technologies, e.g., creating gender-neutral voices for communication that reduces gender bias and language-fluency-controllable TTS for computer-assisted language learning~\cite{KORZEKWA202222}.

In this paper, we propose a method for intermediating multiple speaker attributes by means of optimal-transport-based interpolation of GMMs. Our method first computes the weighted barycenter of a set of GMMs~\cite{agueh2011barycenters}, in which each GMM corresponds to one categorical speaker attribute. Then, it defines a new distribution using the weighted barycenter for sampling nonexistent speakers with a mid-attribute controlled by interpolation weights. One can define such a mid-attribute GMM by estimating its parameters from the interpolation weights representing intermediate speaker attributes. However, this simple method does not guarantee to estimate the best path for interpolating the learned GMMs, since the order of the mixtures in the GMMs is indefinite. In contrast, the optimal transport theory supports the smooth interpolation of multiple distinct distributions and fits our focus on the interpolation of multiple speaker-attribute-aware GMMs.
We empirically validate our method by evaluating the naturalness of synthetic speech and the controllability of two speaker attributes: gender and language fluency. The evaluation results show that our method can control the generated speakers' attributes by a continuous scalar value without statistically significant degradation of speech naturalness. The speech samples are available on our project page\footnote{\scriptsize\url{https://sarulab-speech.github.io/demo_mid-attribute-speaker-generation}}.

\vspace{-3mm}
\section{Related works} \label{sec:relatd-work}
\vspace{-2mm}

\subsection{Speaker embedding prior of TacoSpawn} \vspace{-2mm}
\label{tacospawn}
\drawfig{t}{\linewidth}{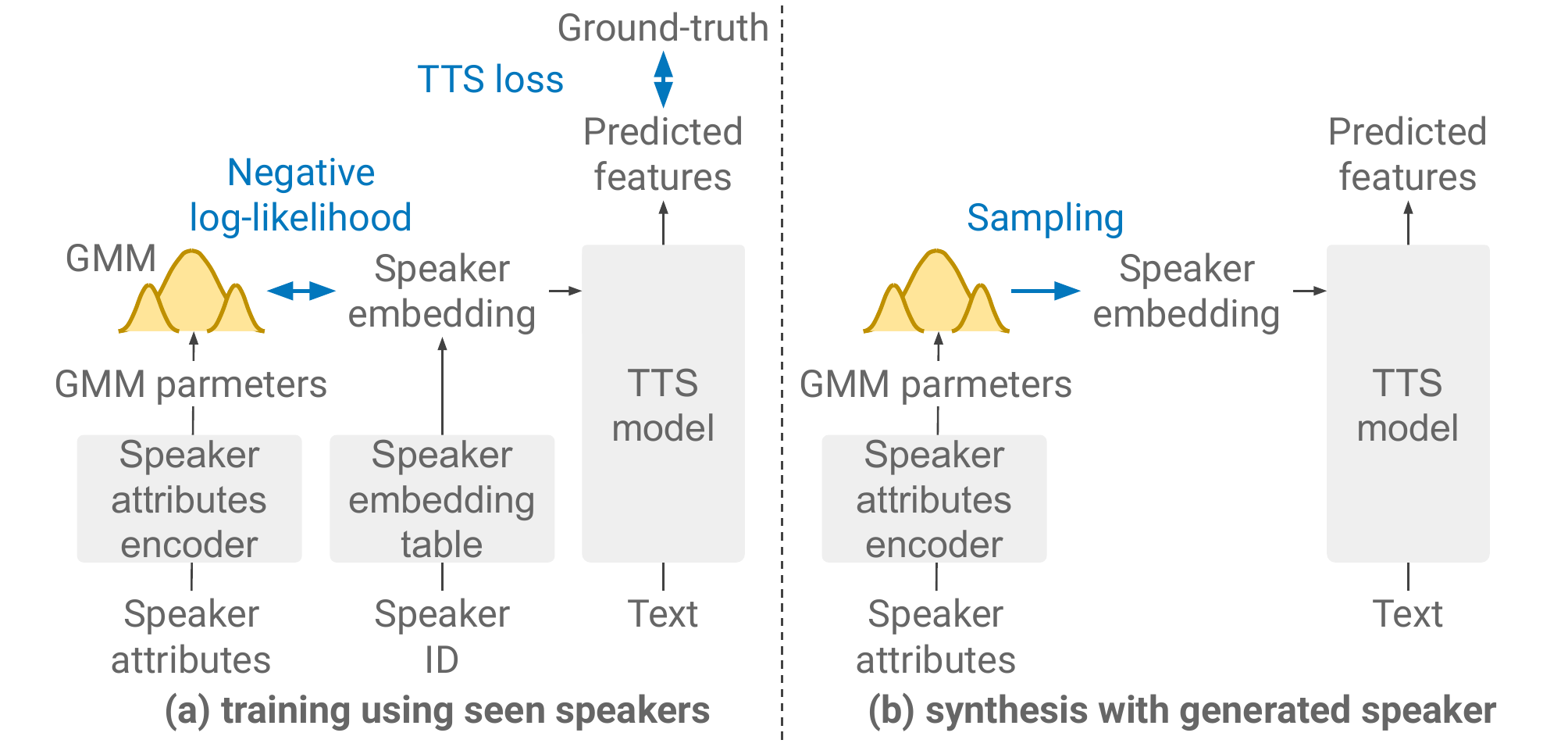}
{Training and synthesis of TacoSpawn.}
TacoSpawn~\cite{stanton2022speaker} learns the speaker embedding distribution as a speaker-attribute-aware GMM and uses the learned distribution to generate a nonexistent speaker's voice. 
Let $D$ and $K$ be the dimensionality of the speaker embeddings and the number of mixture components of the GMM, respectively. A DNN-based encoder takes a speaker attribute $c$ as the input and predicts the GMM parameters, i.e., the non-negative mixture weights $\Vec{\alpha}(c) \in \mathbb{R}^K$, the mean vectors $\Vec{\mu}(c) \in \mathbb{R}^{K\times D}$, and the non-negative variance vectors $\Vec{\sigma}^2(c) \in \mathbb{R}^{K\times D}$ that represent the components of the diagonal covariance matrices. With these parameters, the speaker embedding prior $p\pt{\Vec{s}|c}$ that represents the distribution of the speaker embedding vector $\Vec{s} \in \mathbb{R}^D$ is described as
\begin{align}
    p \pt{\Vec{s} | c} &= \sum _{k=1}^K \alpha_{k}\pt{c} \Normal{\Vec{s}; \Vec{\mu}_k\pt{c}, \Vec{I} \Vec{\sigma}^2_k\pt{c}},\label{eq:pomega}
\end{align}
where $\Normal{\cdot}$ is the Gaussian distribution and $\alpha_{k}\pt{c}$, $\Vec{\mu}_k\pt{c}$, and $\Vec{\sigma}^2_k\pt{c}$ are the $k$th component of $\Vec{\alpha}(c)$, $\Vec{\mu}(c)$, and $\Vec{\sigma}^2(c)$, respectively, and are the parameters of the $k$th mixture. $\Vec{I} \in \mathbb{R}^{D\times D}$ is the identity matrix.

The DNN-based encoder consists of a simple multi-layer perceptron and some activation functions. The objective function for the DNN training is defined as the negative log-likelihood, $-\sum_{j} \log p\left(\Vec{s}_j | c_j\right)$, where $\Vec{s}_j$ and $c_j$ are the $j$th speaker's embedding vector from the speaker embedding table and $j$th speaker's attribute, respectively. Note that this DNN is jointly trained with the TTS model.
\Fig{fig/tacospawn} shows the overview of the TacoSpawn.

\vspace{-2mm}
\subsection{Optimal transport theory} \vspace{-2mm}
The optimal transport is an optimization problem to find the optimal mapping $T\colon \mathbb{R}^n \to \mathbb{R}^n\colon \Vec{x} \mapsto T\left(\Vec{x}\right)$ that moves from a distribution $p_{\rm{a}}$ to a distribution $p_{\rm{b}}$ such that $\int_{\mathbb{R}^n}p_{\rm{a}}\left(\Vec{x}\right)d\Vec{x} = \int_{\mathbb{R}^n}p_{\rm{b}}\left(\Vec{x}\right)d\Vec{x}$.
The optimal mapping in this case minimizes the sum of costs computed by a defined cost function $C\left(\Vec{x}, T\left(\Vec{x}\right)\right)$. In the case of a continuous probability distribution, the objective is expressed by
\begin{align}
    \min_T \int_{\mathbb{R}^n}C\left(\Vec{x}, T\left(\Vec{x}\right)\right)p_{\rm{a}}\pt{\Vec{x}}d\Vec{x},
\end{align}
where
\begin{align}
    \int_{\left\{\Vec{x} \in \mathbb{R}^n; T\pt{\Vec{x}} =  \Vec{y}\right\}}p_{\rm{a}}\pt{\Vec{x}}d\Vec{x} &= p_{\rm{b}}\pt{\Vec{y}}.
\end{align}
This is known to be an ill-posed problem; it is often carried out in practice under conditions that permit transport from one point to several points, in line with Kantorovich's setting~\cite{kantorovich1942translocation}.

The optimal $T\left(\Vec{x}\right)$ guides a transportation path with the minimal cost between two distributions (see \cite{villani2021topics} for the details), and the intermediate state in the path gives the interpolated distribution, which is called a ``barycenter'' in the context of optimal transport. The optimal transport can move a distribution, or a continuous mass, along an optimal path, and can heuristically modify the path with more desirable characteristics by manipulating the cost function. For example, a previous method~\cite{chen2018optimal} defines the movement of a GMM by using the original cost function so as to form the state of the GMM during that movement, which can maintain the same shape before moving.

\vspace{-3mm}
\section{Proposed Method} \label{sec:model_construction}

\drawfig{t}{0.8\linewidth}{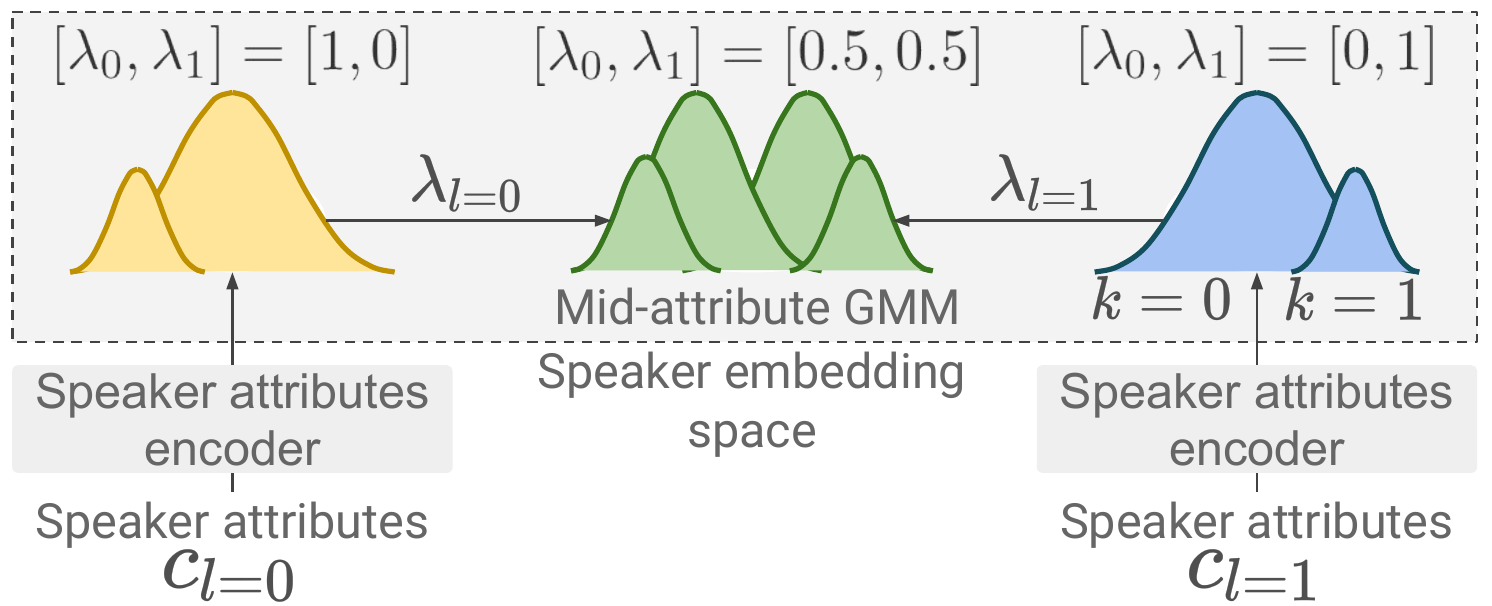}
{Mid-attribute GMM, where $L=2, K=2$.}
\vspace{-2mm}
\subsection{TTS model with speaker attribute encoder}
\vspace{-2mm} \label{TTS model}
We use a multi-speaker TTS model with a TacoSpawn~\cite{stanton2022speaker}-based speaker attribute encoder trained in the same manner as TacoSpawn.
The TTS model, constructed as a sequence-to-sequence model, predicts speech features from a phoneme sequence and a speaker embedding vector $\Vec{s}$.
At the same time, the speaker attribute encoder, constructed with an embedding layer, predicts parameters of a speaker-attribute-aware GMM from the desired speaker attribute $c$. The training and synthesis processes are the same as those in TacoSpawn shown in \Fig{fig/tacospawn}.
Supposing there are $L$ kinds of speaker attributes $\left\{c_l| l=1, 2, \ldots,L\right\}$ , we create in total $L$ $K$-mixture GMM $p_l\pt{\Vec{s}}$ for each $c_l$. The $l$th GMM is defined as:
\begin{align}
    p_l\pt{\Vec{s}} \equiv p\pt{\Vec{s}|c_l} = \sum\limits_{k=1}^{K} \alpha_{l,k} \Normal{\Vec{s}; \Vec{\mu}_{l,k}, \Vec{I}\Vec{\sigma}^2_{l,k}},
\end{align}
where $\alpha_{l,k}, \Vec{\mu}_{l, k}$, and $\Vec{\sigma}_{l, k}$ are the mixture weight, mean vector, and standard-deviation vector of the $k$th mixture of the $l$th GMM, respectively.
From each $p_l\pt{\Vec{s}}$, we generate speaker embeddings with the attribute $c_l$.

\vspace{-2mm}
\subsection{Speaker attributes intermediation} \vspace{-2mm} \label{speaker_distribution_mixing}
We present a method to obtain the intermediate speaker attributes by interpolating $L$ speaker-attribute-aware GMMs $\left\{p_l\left(\Vec{s}\right)|l=1, 2, \ldots, L\right\}$. We define a \textit{mid-attribute GMM} as the weighted barycenter of the $L$ GMMs with interpolation weights $\left\{\lambda_l|l=1, 2, \ldots, L\right\}, \sum_{l} \lambda_l = 1$, as illustrated in \Fig{fig/gmm-optimal-transport}.

\vspace{-2mm}
\subsubsection{Weighted barycenter of non-mixture Gaussian distributions: preliminary} \vspace{-2mm}
First, we describe the simplest case: the weighted barycenter with each distribution being a single Gaussian distribution. To solve this problem as the optimal transport, we define the cost function as a square of Wasserstein distance $W_2$~\cite{villani2021topics, jordan1998variational, otto2001geometry, villani2009optimal} between the two Gaussian distributions $\mathcal{N}\left(\Vec{\mu}_0, \Vec{I} \Vec{\sigma}^2_0\right)$ and $\mathcal{N}\left(\Vec{\mu}_1, \Vec{I} \Vec{\sigma}^2_1\right)$. The cost function is given as~\cite{takatsu2011wasserstein}
\begin{align}
    \label{eq:w2}
 \|\Vec{\mu}_0 - \Vec{\mu}_1\|_2^2 + \|\Vec{\sigma}_0 - \Vec{\sigma}_1\|_2^{2},
\end{align}
where $\|\cdot\|_2^2$ is the $\ell_{2,2}$ norm.

According to \cite{agueh2011barycenters}, the weighted barycenter of $L$ Gaussian distributions $\left\{\mathcal{N}\left(\Vec{\mu}_l, \Vec{I} \Vec{\sigma}_l^2\right)|l=1, 2, \ldots,L\right\}$ 
with interpolation weights $\left\{\lambda_1, \lambda_2, \ldots,\lambda_L\right\}$ is also a Gaussian distribution $\mathcal{N}\left(\Vec{\mu}', \Vec{I} \Vec{\sigma}'^2\right)$. The $\Vec{\mu}'$ and $\Vec{\sigma}'$ are obtained by minimizing the objective function
\begin{align}
    \min_{\Vec{\mu}', \Vec{\sigma}'}\sum_{l=1}^L\lambda_l W_2\pt{\mathcal{N}\pt{\Vec{\mu}', \Vec{I}\Vec{\sigma}'^2}, \mathcal{N}\pt{\Vec{\mu}_l, \Vec{I}\Vec{\sigma}_l^2}}^2. \label{eq:objective_func}
\end{align}
By solving \Eq{objective_func} with \Eq{w2}, they are given as
\begin{align}
    \Vec{\mu}' = \sum_{l=1}^L \lambda_l\Vec{\mu}_l, \quad
    \Vec{\sigma}' = \sum_{l=1}^L \lambda_l \Vec{\sigma_{l}}. \label{eq:sigma_diag}
\end{align}

\vspace{-2mm}
\subsubsection{Weighted barycenter of GMMs} \vspace{-2mm}
The weighted barycenter of $L$ GMMs $\left\{p_l\left(\Vec{s}\right)\right\}$ with interpolation weights  $\left\{\lambda_l\right\}$ is calculated as a combination from $M$ candidates of Gaussian distributions, $\left\{\mathcal{N}\left(\Vec{\mu}'_m, \Vec{I} \Vec{\sigma}'^2_m\right) | m=1, 2, \ldots, M\right\}$,
\begin{align}
    \min_{\Vec{\mu}'_m, \Vec{\sigma}'_m}\sum_{l=1}^L \lambda_l W_2\pt{\mathcal{N}\pt{\Vec{\mu}'_m, \Vec{I}\Vec{\sigma}'^2_m}, p_{l, k_{l, m}}\pt{\Vec{s}}}^2, \label{eq:argmin}
\end{align}
where $m$ is the mixture index of the barycenter and $p_{l, k_{l, m} }\pt{\Vec{s}} = \Normal{\Vec{s}; \Vec{\mu}_{l, k_{l, m}}, \Vec{I} \Vec{\sigma}_{l, k_{l, m}}^2 }$ is the $k_{l,m}$th mixture of $p_l\left(\Vec{s}\right)$ with the mixture weight $\alpha_{l, k_{l, m}}$. 
Each $m$ is assigned to one of possible sequences of $\br{k_l, | k=1, 2, \ldots ,L} \in \left\{1, 2, \ldots ,K\right\}^L$, where $k_l$ is a mixture index of the $l$th GMM.
$k_{l,m}$ indicates $k_l$ assigned to $m$. Thus, the number of mixtures is equal to the number of sequences, $M=K^L$.
The optimal $\Vec{\mu}'_m$ and $\Vec{\sigma}'^2_m$ are obtained as \Eq{sigma_diag}, e.g., $\Vec{\mu}'_m = \sum_{l=1}^{L} \lambda_l \Vec{\mu}_{l, k_{l, m}}$.

The mixture weights $\left\{\alpha'_m\right\}$ for each $\mathcal{N}\left(\Vec{\mu}'_m, \Vec{I} \Vec{\sigma}'^2_m\right)$ 
are estimated by solving
\begin{align}
    \min_{\pi_{l,k,m}} \sum\limits_{l=1}^{L}\sum\limits_{k =1}^{K}\sum\limits_{m=1}^{M} 
        \lambda_l \pi_{l,k,m} W_2\pt{\Normal{\Vec{\mu}'_m, \Vec{I} \Vec{\sigma}_m'^2}, p_{l, k}\pt{\Vec{s}}}^2, \label{eq:minimization_bary}
\end{align}
under
\begin{align}
    \sum\limits_{m=1}^{M} \pi_{l,k,m} = \alpha_{l, k}, \;\;\;\;
    \sum\limits_{k=1}^{K} \pi_{1,k,m} 
    = \cdots = \sum\limits_{k=1}^{K} \pi_{L,k,m}. \label{eq:const_kdash}
\end{align}
Solving \Eq{minimization_bary}, $\alpha'_m$ is given as
\begin{align}
    \alpha_m' = \sum\limits_{k=1}^{K} \pi_{1,k,m}.
\end{align}

Minimization of \Eq{minimization_bary} can be regarded as the optimal transport of discrete probability distributions.
In our implementation, we ignore the right side of \Eq{const_kdash} to simplify the optimization of $\pi_{l, k, m}$. The optimal transport thus becomes a hard mapping: rather than mapping all mixtures of each GMM, one sub-optimal (i.e., closest in the square of the Wasserstein distance) mixture in each GMM is mapped to one closest mixture of the barycenter. Namely, $\pi_{l, k, m}$ is set to $\lambda_l \alpha_{l, k}$ for the sub-optimal mixture and $0$ for the others.

\subsubsection{Speaker sampling of intermediated attribute} \vspace{-1mm}
We now describe how to generate new speakers with mid-attributes and synthesize their voices using the models described in Section \ref{TTS model}.
To generate a new speaker with a mid-attribute, we randomly sample a vector from the speaker-attribute-aware GMM of the target mid-attribute. Using this sampled vector, we can synthesize the voice with the characteristics of the target mid-attribute.
For example, if $c_1$ means Japanese male and $c_2$ is Japanese female, the weighted barycenter of $\left\{p_l\left(\Vec{s}\right)\right\}$ with interpolation weights $\left\{\lambda_{l = 1, 2} = 0.5, \lambda_{3 \leq l \leq L} = 0 \right\}$ represents the middle attribute between Japanese male and Japanese female, which can be regarded as a Japanese speaker with a mid-attribute in gender.

\vspace{-2mm}
\section{Experimental evaluation} \label{sec:experiment}
\vspace{-2mm}
\subsection{Experimental setup}\label{subsec:condition} \vspace{-1mm}
In this experiment, we built a speaker generation model for Japanese TTS that can control the speaker's gender and language fluency (i.e., nativeness $=$ native-accented through foreign-accented). 
As training corpora, we utilized JVS~\cite{takamichi2020jsut} (Japanese) and VCTK~\cite{veaux2017vctk} (English), which include $100$ Japanese speakers ($49$ males and $51$ females) and $108$ English speakers ($47$ males and $61$ females), respectively. 
The attribute of each speaker was described as a combination of the speaker's gender and nativeness. Since we aimed to synthesize Japanese speech, English speakers were considered as ``non-native speaker'' (this nativeness-aware training is described below). For example,  the attribute of a male speaker in JVS is set to [male, native], and a female speaker in VCTK is set to  [female, non-native]. Thus, there were $L=4$ kinds of speaker attributes available for this experiment. 
Additionally, we used JSUT~\cite{takamichi2020jsut}, a Japanese single-speaker corpus, for pre-training the TTS model.
All speech data were resampled to 22.05 kHz.

Our expectation is that intermediating the attributes can not only generate a gender-neutral speaker's voice but also control the nativeness of the speaker. However, our preliminary experiments had shown that naive implementation didn't reflect the nativeness.
To achieve the nativeness-aware TTS, we trained a multi-lingual TTS model with a long short-term memory (LSTM)~\cite{hochreiter1997long}-based native-language classifier.  The classifier estimates the speaker's native language from a mel-spectrogram generated by the TTS model. At the same time, the TTS model is trained to make the native-language classification easier. The DNN architecture for the classifier was based on Xin et al.'s work~\cite{xin21_interspeech}. We empirically confirmed that the TTS model synthesized non-native speech reflecting the characteristics of native language, i.e., English-accented Japanese speech.

We used a FastSpeech 2~\cite{ren2020fastspeech}-based TTS model.  The TTS training objective here was the same as that in original FastSpeech 2. We used a publicly available PyTorch implementation\footnote{\scriptsize\url{https://github.com/Wataru-Nakata/FastSpeech2-JSUT}} of FastSpeech 2 followed by a pre-trained HiFi-GAN~\cite{kong2020hifi} neural vocoder\footnote{\scriptsize\url{https://github.com/jik876/hifi-gan}}.
Input texts were phonemized with pyopenjtalk\footnote{\scriptsize\url{https://github.com/r9y9/pyopenjtalk}} for Japanese and eSpeak NG\footnote{\scriptsize\url{https://github.com/espeak-ng/espeak-ng}} for English.

\drawfig{t}{0.98\linewidth}{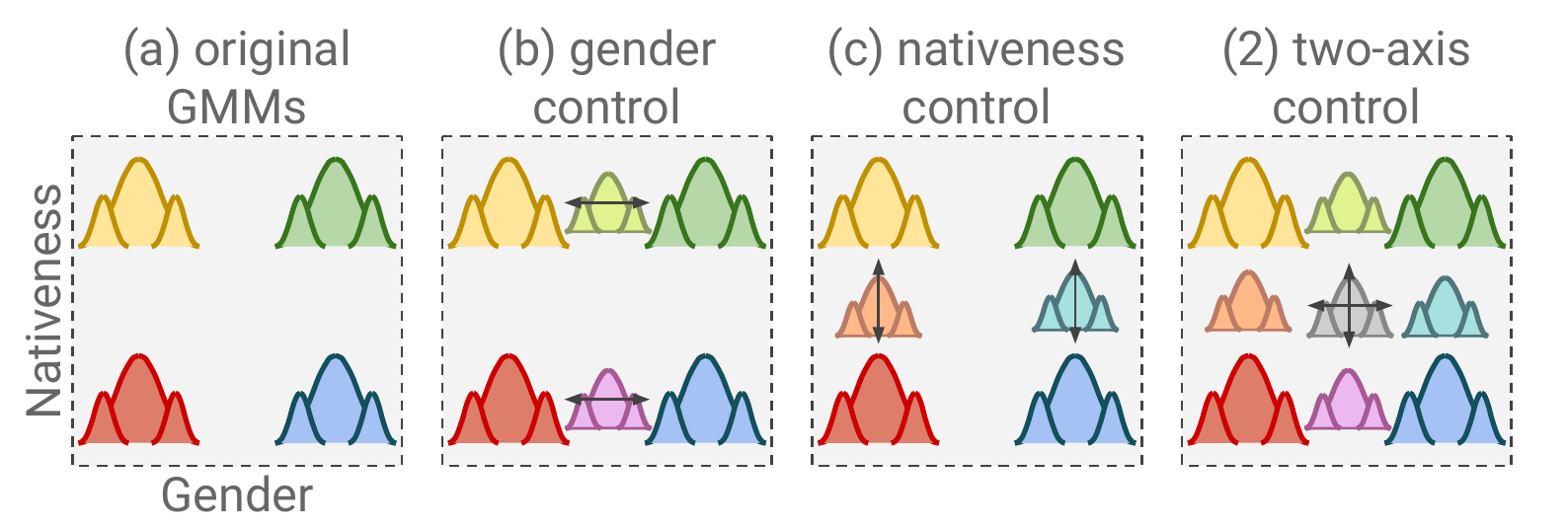}
{Target distributions in the experiment.}

For speaker generation, we used $3$-mixture GMMs as the speaker distribution that modeled $256$-dimensional speaker embeddings.
As shown in \Fig{fig/mixing_in_experiment.pdf}(a), our model learned four GMMs representing the four speaker attributes along two axes: gender and nativeness. We then used these GMMs in two experiments: 1) one-axis control and 2) two-axis control. The former interpolated the two GMMs of two attributes by changing one axis and fixing the other, as shown in \Fig{fig/mixing_in_experiment.pdf}(b) and (c). Namely, the interpolation weights took 0.5 for the two used GMMs and 0.0 for the others. The latter investigated the interaction between the two axes by interpolating the four mid-attribute GMMs, as shown in \Fig{fig/mixing_in_experiment.pdf}(d). Namely, all the interpolation weights were 0.25. In summary, we used nine distributions for this evaluation, i.e., four categorical-attribute GMMs and five mid-attribute GMMs. Actually, we conducted evaluations at even finer intervals 
(e.g., weights of $0.25$ and $0.75$ in the one-axis control)
, but we show only the results of the above nine GMMs due to space constraints.
In addition, we did not evaluate existent speakers' synthetic speech samples, since Stanton et al. had already demonstrated that TacoSpawn could synthesize voices of nonexistent speakers with categorical attributes as natural as them~\cite{stanton2022speaker}. 
\vspace{-2mm}
\subsection{Subjective evaluation} \vspace{-2mm}
We carried out the subjective evaluation of our method from three perspectives: perceived gender, perceived nativeness, and naturalness of synthetic speech. We synthesized the Japanese speech of 225 (25 $\times$ 9 distributions) nonexistent speakers generated by our model.
Sentences of the recitation324 subset in the ITA corpus\footnote{\scriptsize\url{https://github.com/mmorise/ita-corpus}} were used for the evaluation.
We conducted the subjective evaluation using Lancers\footnote{\scriptsize\url{https://www.lancers.jp/}}, a Japanese crowdsourcing platform. We recruited $750$ listeners for each evaluation. Each listener evaluated $17$ speech samples with a 5-scale integer. For convenience of numerical evaluation, we assigned female and male to $-2$ and $+2$ on the ``Gender'' axis, respectively, and the sign of the number had no particular meaning. The naturalness rating ranged from 1 (very unnatural) to 5 (very natural). The speaker attributes were balanced among the samples described in Section \ref{subsec:condition}. The linguistic contents of the sentences we used did not evoke gender or nativeness, but even so, we instructed the listeners not to pay attention to the content of the presented utterances when evaluating the speech samples. At least $20$ listeners evaluated one speaker's synthetic speech. The listeners' answers were aggregated for each speaker. In the nativeness evaluation, we presented English-accented Japanese natural speech from the UME-JRF corpus~\cite{umejrf} to the listeners as an example of non-native speech.

\begin{figure}[t]
    \centering
    \begin{tabular}{c}
        \begin{minipage}{0.48\linewidth}
            \centering
            \includegraphics[width=\linewidth]{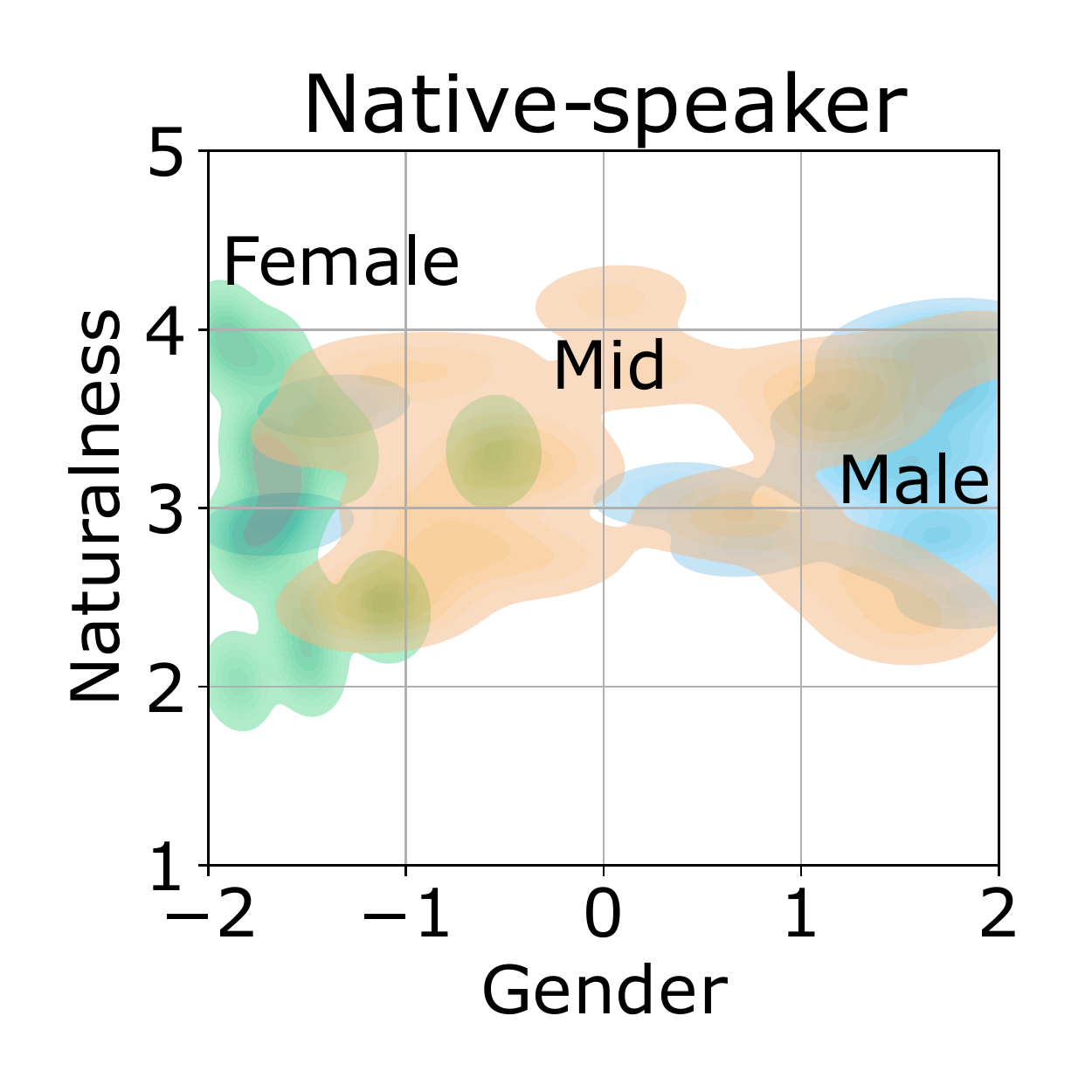}
            \vspace{-10mm}
            \caption{Gender-axis control within native speakers}
            \label{fig:fig/takamichi/Gender-control_in_Native-speaker}
        \end{minipage}
        \begin{minipage}{0.48\linewidth}
            \centering
            \includegraphics[width=\linewidth]{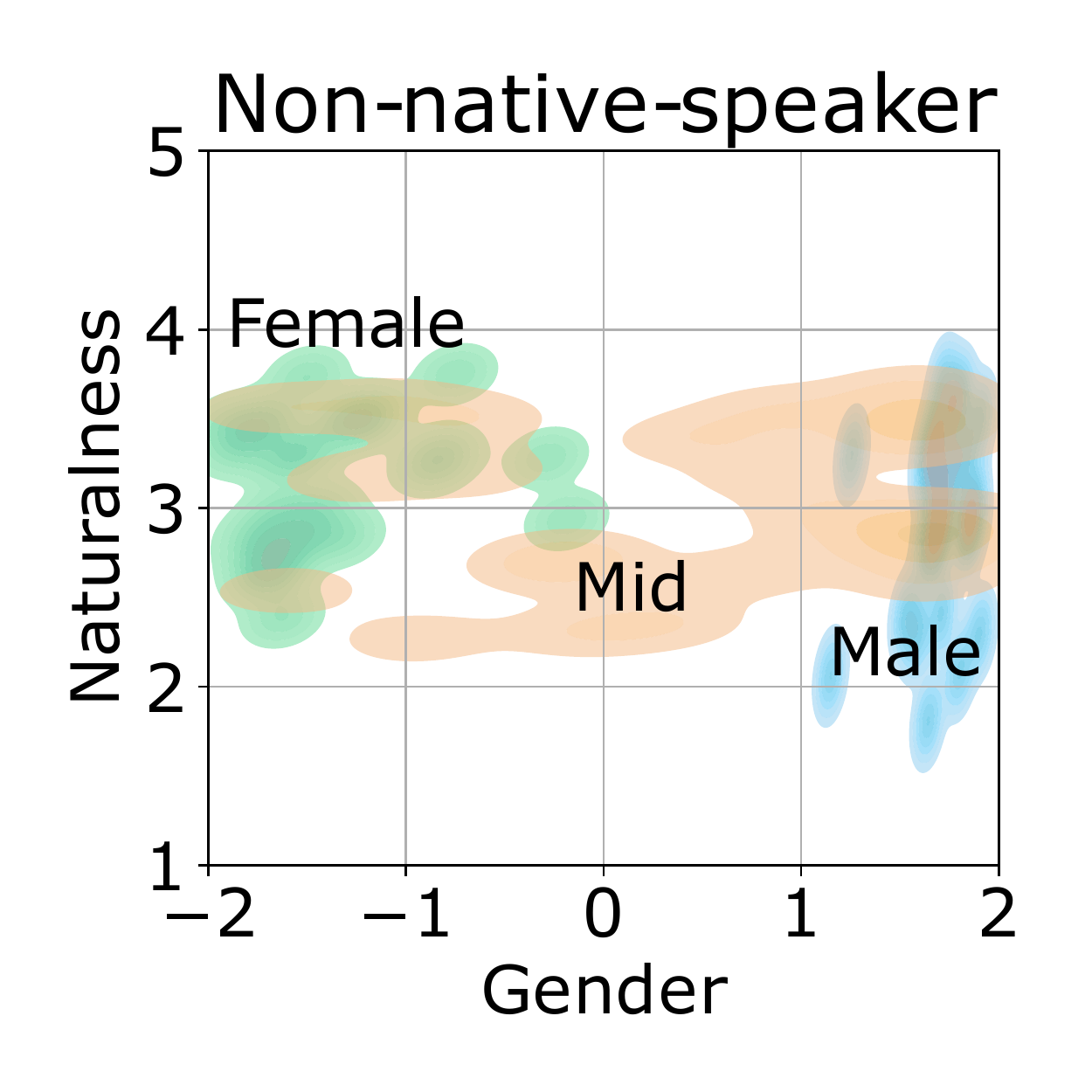}
            \vspace{-10mm}
            \caption{Gender-axis control within non-native speakers}
            \label{fig:fig/takamichi/Gender-control_in_Non-native-speaker}
        \end{minipage}
    \end{tabular}
    \vspace{-5mm}
\end{figure}
\begin{figure}[t]
    \centering
    \begin{tabular}{c}
        \begin{minipage}{0.48\linewidth}
            \centering
            \includegraphics[width=\linewidth]{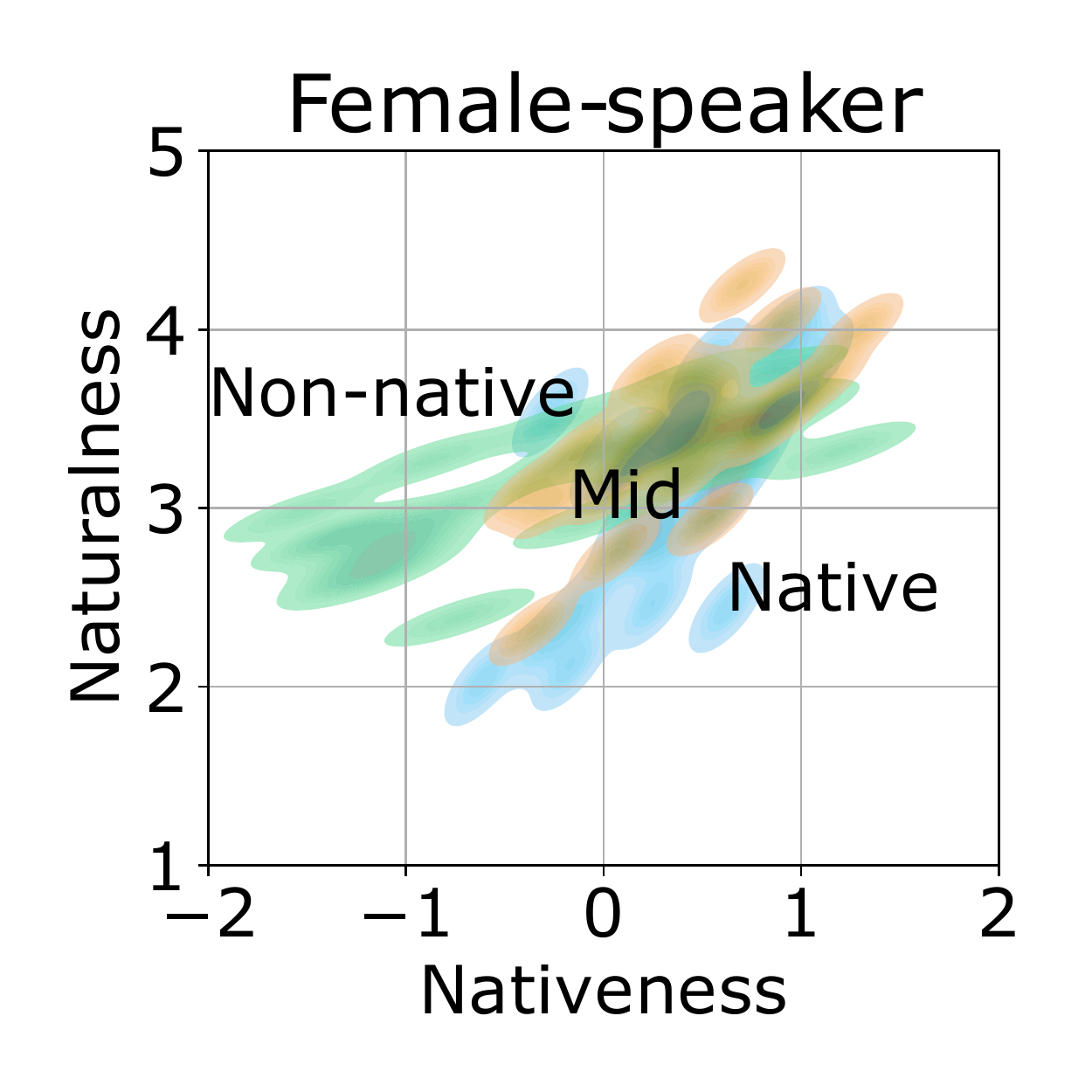}
            \vspace{-10mm}
            \caption{Nativeness-axis control within female speakers}
            \label{fig:ffig/takamichi/Nativeness-control_in_Female-speaker}
        \end{minipage}
        \begin{minipage}{0.48\linewidth}
            \centering
            \includegraphics[width=\linewidth]{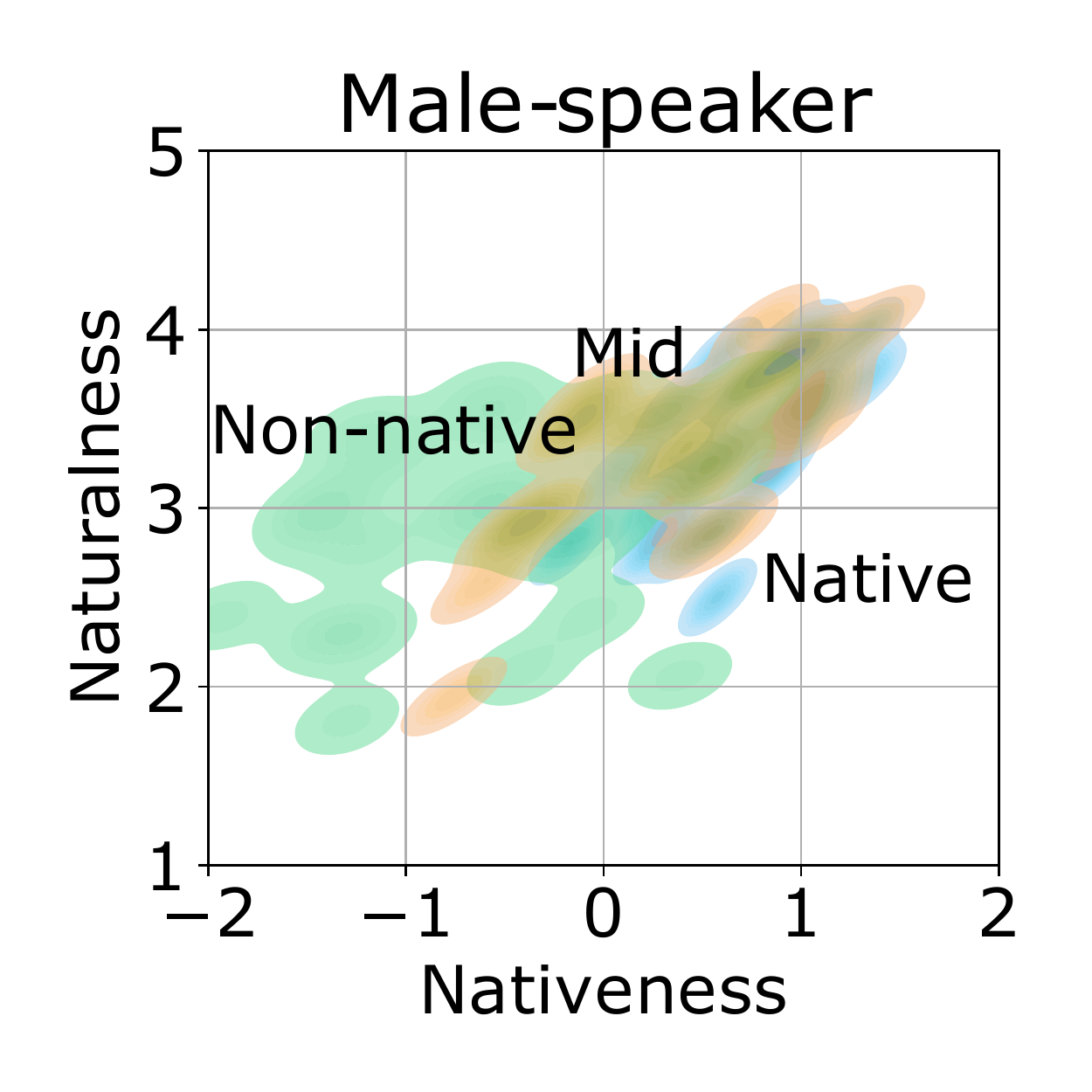}
            \vspace{-10mm}
            \caption{Nativeness-axis control within male speakers}
            \label{fig:ffig/takamichi/Nativeness-control_in_Male-speaker}
        \end{minipage}
    \end{tabular}
    \vspace{-5mm}
\end{figure}

\begin{figure}[t]
    \centering
    \begin{tabular}{c}
        \begin{minipage}{0.48\linewidth}
            \centering
            \includegraphics[width=\linewidth]{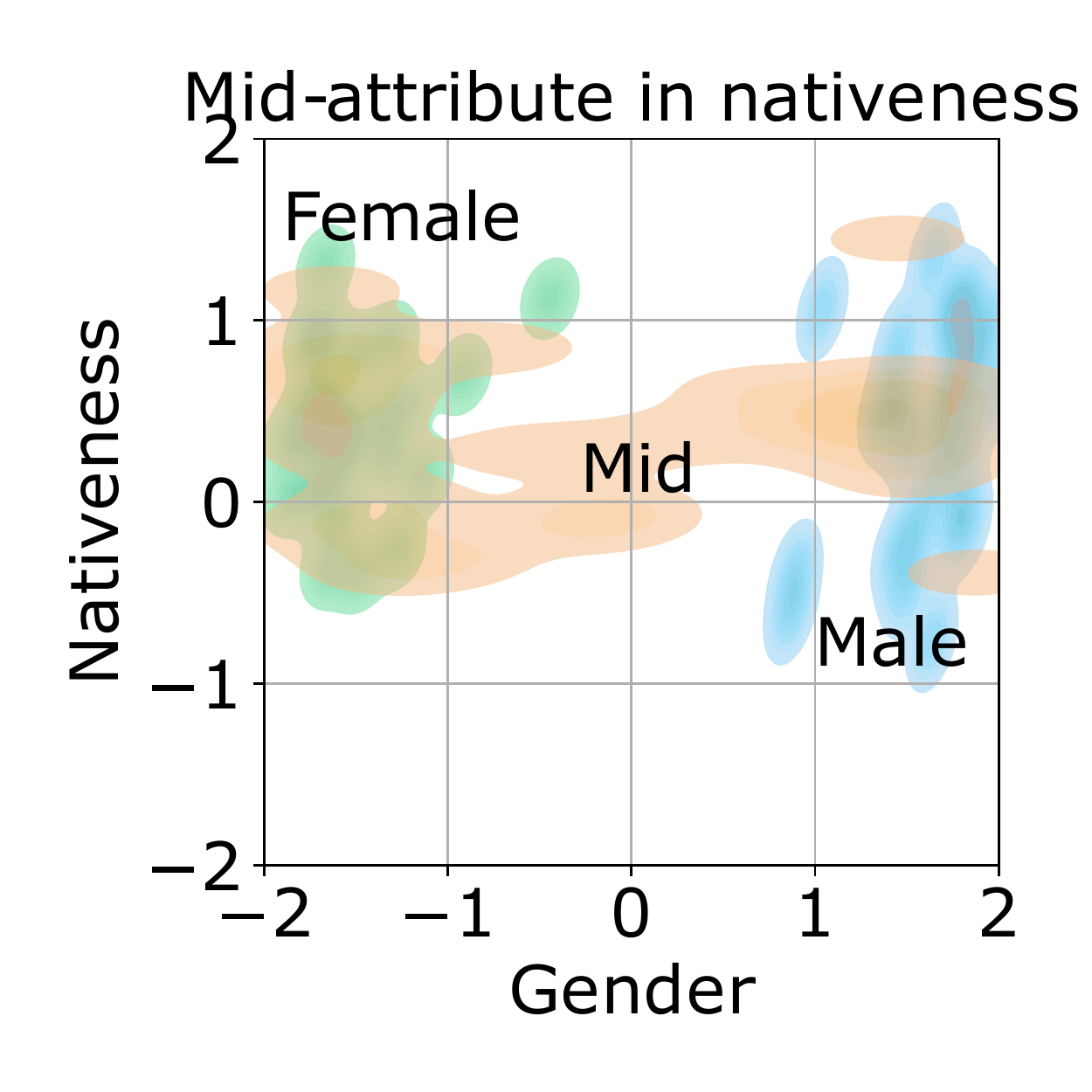}
            \vspace{-10mm}
            \caption{Gender-axis control within mid-attribute speakers in nativeness.}
            \label{fig:ffig/takamichi/Gender-control_in_Mid-native-speaker}
        \end{minipage}
        \begin{minipage}{0.48\linewidth}
            \centering
            \includegraphics[width=\linewidth]{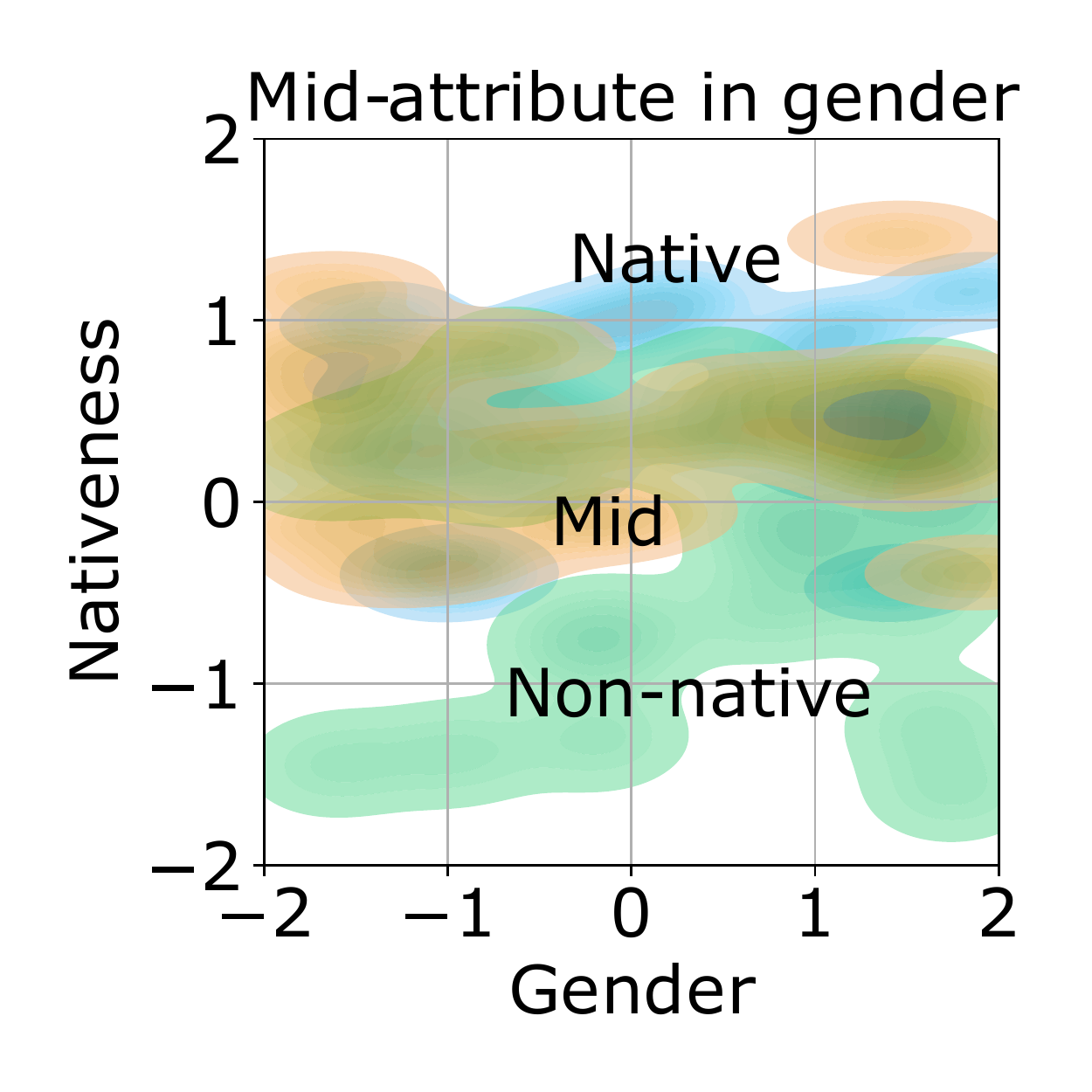}
            \vspace{-10mm}
            \caption{Nativeness-axis control within mid-attribute speakers in gender.}
            \label{fig:ffig/takamichi/Nativeness-control_in_Mid-gender-speaker}
        \end{minipage}
    \end{tabular}
    \vspace{-3mm}
\end{figure}
We visualized the evaluation results by kernel density estimation (KDE) plots.
Figure \ref{fig:fig/takamichi/Gender-control_in_Native-speaker} and Figure \ref{fig:fig/takamichi/Gender-control_in_Non-native-speaker} show the KDE plots of the perceived gender and naturalness of synthetic speech within native and non-native speakers, respectively. ``Male'' and ``Female'' are generated by the TacoSpawn framework, and ``Mid'' is generated by our method using the optimal-transport-based interpolation. We can see that the perceived gender of synthetic speech with the mid-attribute (``Mid'') is widely distributed while keeping the naturalness achieved by the original speaker generation with categorical attributes, i.e., ``Male'' and ``Female.'' This result indicates that our proposed method can cover wider speaker distributions for speaker generation comparing to the conventional TacoSpawn.
We observe a similar tendency in the KDE plots of perceived nativeness ($-2$: non-native to $+2$: native) and naturalness of synthetic speech within female and male speakers shown in Figure \ref{fig:ffig/takamichi/Nativeness-control_in_Female-speaker} and Figure \ref{fig:ffig/takamichi/Nativeness-control_in_Male-speaker}, respectively. One noteworthy point here is that the naturalness tends to increase in proportion to the nativeness. This result is quite understandable because the TTS model never observes the ground-truth non-native Japanese speech data during the training. We also find that the boundary between the two distributions of non-native and native speakers is more ambiguous than those shown in Figure \ref{fig:fig/takamichi/Gender-control_in_Native-speaker} and Figure \ref{fig:fig/takamichi/Gender-control_in_Non-native-speaker}. One reason might be the characteristics of the evaluation data, which included some sentences composed of rare moras and strange words in Japanese.

Figure \ref{fig:ffig/takamichi/Gender-control_in_Mid-native-speaker} and Figure \ref{fig:ffig/takamichi/Nativeness-control_in_Mid-gender-speaker} show 
the results of two-axis control experiments, i.e., gender-axis control within mid-attribute speakers in nativeness and nativeness-axis control within mid-attribute speakers in gender, respectively. We can see that the ``Mid'' distributions in these two figures cover the middle ranges of both gender and nativeness. From these results, we conclude that our method can interpolate even the mid-attribute distributions and achieve the cross-attribute control of mid-attribute speakers.

\begin{figure}[t]
    \centering 
    \vspace{0mm}
    \includegraphics[width=\linewidth]{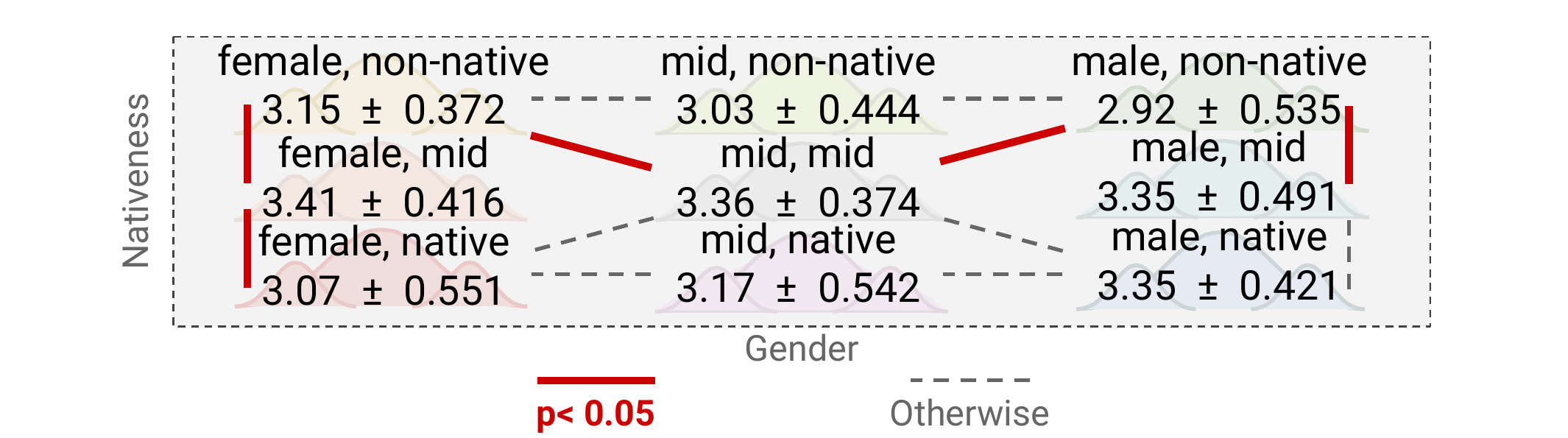} 
    \vspace{-8.0mm}
    \caption{Mean and standard deviation of naturalness for each attribute speaker. Red line denotes the significant differences between the two connected scores.}
    \label{fig:fig/mos_p.pdf}
    \vspace{-4mm}
\end{figure}

\Fig{fig/mos_p.pdf} shows mean opinion scores (MOS) values regarding the naturalness of speech generated by using each of the nine distributions. We observe that MOS values of mid-attribute speakers' synthetic speech are comparable to or higher than those of synthetic speech with the categorical attributes. This result demonstrates that our method achieves mid-attribute speaker generation without degrading the naturalness of synthetic speech.

\vspace{-2mm}
\section{Conclusion} \label{sec:conclusion}
\vspace{-2mm}
In this paper, we proposed a method for generating nonexistent speakers with mid-attributes by means of the optimal-transport-based interpolation of speaker-attribute-aware GMMs. Subjective experiments confirmed that our method can generate nonexistent speakers that have perceptible mid-attributes, without significantly degrading speech naturalness.
We plan to increase the language diversity in the future. 

\newpage
\printbibliography


\end{document}